\documentclass[aps,11pt]{revtex4}
\usepackage{dcolumn}
\usepackage{graphicx}
\usepackage{amsmath}
\usepackage{amsfonts}
\usepackage{amssymb}
\usepackage{psfrag}
\usepackage{wrapfig}
\usepackage{subfigure}
\usepackage{makeidx}
\usepackage{bm}
\usepackage{epsf}
\usepackage{hyperref}
 \usepackage{float}
\usepackage{color}
\usepackage{cases}
\usepackage{booktabs} 
\usepackage{makecell} 
\usepackage{multirow}

\makeatletter

\newcommand{\Rmnum}[1]{\uppercase\expandafter{\romannumeral #1}}
\usepackage{amsmath} 
\usepackage[utf8]{inputenc}  
\usepackage[T1]{fontenc}

\newcolumntype{M}[1]{>{\centering\arraybackslash$}m{#1}<{$}}  

\usepackage{amsmath} 

\usepackage{booktabs}

\begin{document} 

\title{Spin-charge induced scalarization of  Kerr-Newman black holes in the Einstein-Maxwell-scalar theory with scalar potential}

\author{Xiang Luo$^{1}$, Meng-Yun Lai$^{1}$\footnote{mengyunlai@jxnu.edu.cn},Yun Soo Myung$^{2}$\footnote{ysmyung@inje.ac.kr}, Yi-Bin Huang$^{1}$\\ and De-Cheng Zou$^{1}$\footnote{corresponding author:dczou@jxnu.edu.cn}}

\affiliation{$^{1}$College of Physics and Communication Electronics, Jiangxi Normal University, Nanchang 330022, China\\
$^{2}$Center for Quantum Spacetime, Sogang University, Seoul 04107, Republic of Korea}

\vspace{2mm}

\begin{abstract}

We investigate  the spin-charge-induced scalarization of Kerr--Newman (KN) black holes in the Einstein--Maxwell-scalar (EMS) theory with a scalar potential and positive coupling parameter.
In the linearized theory,  there exists  a bound of $0<a<a_o$ with onset  spin $a_c$ for  the negative region signaling  instability  by analyzing the effective scalar mass term in the $\theta$-direction.
Solving the $(2+1)$-dimensional  evolution equation numerically, we find  the region where the KN black hole becomes unstable, 
giving rise to  scalarized KN black holes. The threshold curve for representing the boundary between stable and unstable KN black holes depends on charge $Q$, scalar mass $m_\phi$, coupling parameter $\alpha$, and spin parameter
 $a$ with upper bound $a^2\le M^2-Q^2$. 

\end{abstract}

\maketitle

\section{Introduction}\label{1s}

{{The no-hair theorem} } has long shaped black hole physics in general relativity, stating that mass ($M$), charge 
($Q$), and spin ($a=J/M$) with $J$ angular momentum   describe Kerr--Newman (KN)  black holes~\cite{Carter:1971zc,Ruffini:1971bza}. It rules out regular scalar solutions in flat spacetimes because scalar fields become unstable and blow up at the horizon~\cite{Bekenstein:1974sf,Bekenstein:1975ts,Bronnikov:1978mx}.
In 1993, however,  Damour and Esposito-Farèse found a new way of avoiding  this theorem by  making use of  scalar--tensor theories with special scalar couplings. They showed that scalar hair could survive on black holes without breaking physics~\cite{Damour:1993hw}.
The breakthrough came out when theorists probed two specific models. In scalar--tensor theories that include  the nonminimal scalar coupling to either
the Gauss--Bonnet (GB) term~\cite{Doneva:2017bvd,Silva:2017uqg,Antoniou:2017acq}
or Maxwell term~\cite{Herdeiro:2018wub,Fernandes:2019rez,Herdeiro:2019yjy,Myung:2018vug}, the scalar field triggered destabilization of bald  (scalar-free) black holes
and induced scalarized (charged) black holes. That is, the tachyonic scalar played  an important role in 
triggering  spontaneous scalarization  of the  bald black holes.

Recent works have shown that in scalar-Gauss--Bonnet gravity, scalarization of Kerr black holes depends on spin $a$  and coupling parameter. For a negative coupling parameter, scalarization is allowed  when spin exceeds $a\ge 0.5M$, as confirmed by studies~\cite{Hod:2020jjy,Zhang:2020pko,Doneva:2020nbb,Berti:2020kgk}. In this direction, high rotation triggered  tachyonic instability~\cite{Dima:2020yac}, and it was validated numerically~
\cite{Cunha:2019dwb,Collodel:2019kkx,Herdeiro:2020wei}.  Analyses using different coupling functions~\cite{Zou:2021ybk,Doneva:2022yqu} further confirmed  that scalarization depends on both spin and  coupling parameter.  We remind  the reader that the  spin-induced scalarization of a Kerr black hole with a massive scalar  were also discussed in scalar-Gauss–Bonnet theory~\cite{Doneva:2020kfv} and dynamical Chern–-Simons gravity ~\cite{Zhang:2021btn}.

In the Einstein--Maxwell-scalar (EMs) theory with scalar coupling parameter $\alpha$, scalarization for KN black holes has also been investigated. The threshold curves [$\alpha(a)$] were found with a different charge $Q$,
describing the boundary between  KN  and  scalarized KN black holes in the EMS theory.  For negative coupling parameter, a lower bound of  $\frac{a}{r_+}\geq \hat{a}(=0.4142) $ was found  in the limit of $\alpha\rightarrow-\infty$ by using analytical~\cite{Hod:2022txa} {{and two numerical}~\cite{Lai:2022spn} methods---the hyperboloidal foliation time evolution method~\cite{Zenginoglu:2007jw} and the direct 2+1 time evolution method.}   For the positive coupling parameter, however, there was  no lower bound on $a$, while the upper bound of $a$ ($a^2\le M^2-Q^2$) appeared as the existence condition for the outer horizon~\cite{Lai:2022ppn}.
This  implies   that the high rotation enhances  scalarization for KN black holes in EMs theory.

In the present work, we wish to study the spin-charge-induced scalarization of a KN black hole when the scalar field is massive and the scalar coupling parameter is positive $\alpha>0$. We will adopt the direct 2 + 1 time evolution method \cite{Doneva:2020kfv,Zhang:2021btn} to perform a time evolution of linearized massive scalar  perturbation on the KN black hole background. In this case, we expect to find that  the scalar mass term might  alter tachyonic instability and spin-charge-induced scalarization significantly.

We organize the present work  as follows. In Section~\ref{2s}, we revisit   the linearized scalar theory by analyzing the effective scalar  mass term $\mu^2_{\rm eff}$. We perform the spin-charge-induced scalarization of KN black holes in Section~\ref{3s} by solving the (2 + 1)-dimensional evolution equation numerically and then show the numerical results for different parameters $Q$, $\alpha$, $a$, etc.
, in Section~\ref{4s}.  Section~\ref{5s} is devoted to contributing to conclusions and discussions.


\section{Linearized Scalar Theory}\label{2s}

The action of EMS theory with scalar potential takes the form 
~\cite{Herdeiro:2018wub,Zou:2019bpt}
\begin{eqnarray}
S_{\rm EMS}=\frac{1}{16\pi}\int d^4x{\sqrt{-g}\Big[R-2\partial_\mu\phi\partial^\mu\phi-U(\phi)-f(\phi)F^2\Big]}, \label{action}
\end{eqnarray}
where $R$ is the Ricci scalar, $\phi$ is the scalar field with a potential $U(\phi)$, and  the scalar coupling function $f(\phi)$ is coupled 
to the Maxwell term $ F^2=F_{\mu\nu}F^{\mu\nu}$ with $F_{\mu\nu}=\partial_{\mu}A_{\nu}-\partial_{\nu}A_{\mu}$.

The variation of action \eqref{action} with respect to the metric $g_{\mu\nu}$, scalar field $\phi$, and vector potential $A_\mu$ gives the following  equations:
\begin{eqnarray}
&&R_{\mu\nu}-\frac{1}{2}R g_{\mu\nu}=2\partial _\mu \phi\partial _\nu \phi-(\partial \phi)^2g_{\mu\nu}
+2f(\phi)\Big(F_{\mu\rho}F_{\nu}~^\rho-\frac{F^2}{4}g_{\mu\nu}\Big), \label{g-eql}\\
&&\nabla_{\mu}\nabla^{\mu}\phi -\frac{1}{4}f'(\phi)F^2-\frac{1}{4}U'(\phi)=0, \label{s-equa1}\\
&&\partial_\mu\left(\sqrt{-g}f(\phi)F^{\mu\nu}\right)=0\label{M-eq1},
\end{eqnarray}
where the prime ($'$) denotes the derivative with respect to its argument.

Without scalar hair, the axisymmetric KN black hole solution is expressed in terms of the Boyer--Lindquist coordinates as
\begin{eqnarray}
ds^2_{\rm KN} &\equiv& \bar{g}_{\mu\nu}dx^\mu dx^\nu=-\frac{\Delta-a^2\sin^2\theta}{\rho^2}dt^2
-\frac{2a\sin^2\theta(r^2+a^2-\Delta)}{\rho^2}dt d\varphi\nonumber\\
&&+\frac{[(r^2+a^2)^2-\Delta a^2 \sin^2\theta]\sin^2\theta}{\rho^2} d\varphi^2+ \frac{\rho^2}{\Delta}dr^2 +\rho^2 d\theta^2, \label{KN-sol}
\end{eqnarray}
where
\begin{eqnarray}
 \Delta= r^2-2Mr+a^2+Q^2,\quad \rho^2=r^2+a^2\cos^2\theta.\nonumber
\end{eqnarray}
The corresponding vector potential is given by
\begin{eqnarray}
A=-\frac{Qr}{\rho^2}\left(dt-a\sin^2\theta d\varphi\right).\label{vecp}
\end{eqnarray}
The outer and inner horizons are obtained by imposing  $\Delta=(r-r_+)(r-r_-)=0$ as
\begin{eqnarray}
r_{\pm}=M\pm \sqrt{M^2-a^2-Q^2},
\end{eqnarray}
where one requires the existence condition for two horizons with $M\ge Q$
\begin{equation}
    a^2\le M^2-Q^2 \label{l-b}
\end{equation}
For simplicity, we set the mass of the KN black hole to be $M=1$ in the whole article.

To handle  the tachyonic instability of KN black holes in the EMS theory, we consider the linearized  scalar equation
\begin{eqnarray}
\left(\bar{\square}-\mu_{\rm eff}^2\right)\delta\phi=0, \quad
\mu_{\rm eff}^2=\frac{1}{4}\Big[\bar{F}^2f''(0)+U''(0)\Big]\label{per-eq}.
\end{eqnarray}
This instability is characterized by the presence of a
negative mass term ($\mu^2_{\rm eff}<0)$ in the linearized scalar equation.
Until now, many authors adopted different forms of the coupling function
$f(\phi)$ with coupling parameter $\alpha$,  exponential form ($e^{\alpha\phi^2}$)~\cite{Herdeiro:2018wub},
quadratic form ($1+\alpha\phi^2$)~\cite{Myung:2018vug}, and hyperbolic cosine 
form ($\cosh[\sqrt{2\alpha}\phi]$)~\cite{Fernandes:2019rez}.  We introduce  an interesting  {{potential function  as}  $U(\phi)$= $m_{\phi}^2$$\phi^2 $ with $m^2_\phi$ mass of the scalar, as shown in Ref.~\cite{Zou:2019bpt}.}
Without loss of generality,
we do not choose any specific  form of coupling function $f(\phi)$ but  
require that $f(\phi)$ satisfies~\cite{Fernandes:2019rez,Herdeiro:2019yjy}
\begin{eqnarray}\label{fphi}
f(0)=1,\quad f'(0)=0, \quad f''(0)=2\alpha, \label{fphi2}
\end{eqnarray}
{where $f(0)=1$ implies the presence of the Maxwell term even for non-scalar coupling,  $f'(0)=0$
denotes the presence of a scalar coupling function, and the last $f''(0)=2\alpha$ means the presence of a quadratic scalar function with coupling parameter $\alpha$. }
For the KN black hole background, 
the effective mass squared in Equation~\eqref{per-eq} is given by 
\begin{eqnarray}
\mu_{\rm eff}^2=\frac{\alpha \bar{F}^2}{2}+\frac{1}{2}m_{\phi}^2=-\frac{\alpha Q^2(r^4-6a^2r^2\cos^2\theta+a^4\cos^4\theta)}{\left(r^2+a^2\cos^2\theta\right)^4}+\frac{1}{2}m_{\phi}^2.\label{effmass}
\end{eqnarray}

Intuitively, the influence of $\mu^2_{\rm eff}$ on the tachyonic instability
 could be found  in the near-horizon if one chooses $\theta$  appropriately. 
 For $\theta=\pi/2$, its spin ($a$) dependence disappears, but its contribution is dominant. Hence, we may choose  $\theta=0,\frac{\pi}{3},\frac{0.9 \pi}{2}$ to obtain its simple spin dependence.  In this case, we find the bound for scalarization with positive $\alpha$ as
\begin{equation} \label{a-c}
    0<a<a_o(Q,m_\phi,\alpha),
\end{equation}
where $a_o$ denotes the  onset spin, depending on three parameters of $Q,m_\phi,$ and $\alpha$.

\begin{figure}[h!]
\subfigure[~ $Q=0.6$,~$\theta=0$ ]{  \includegraphics[width=0.3\textwidth]{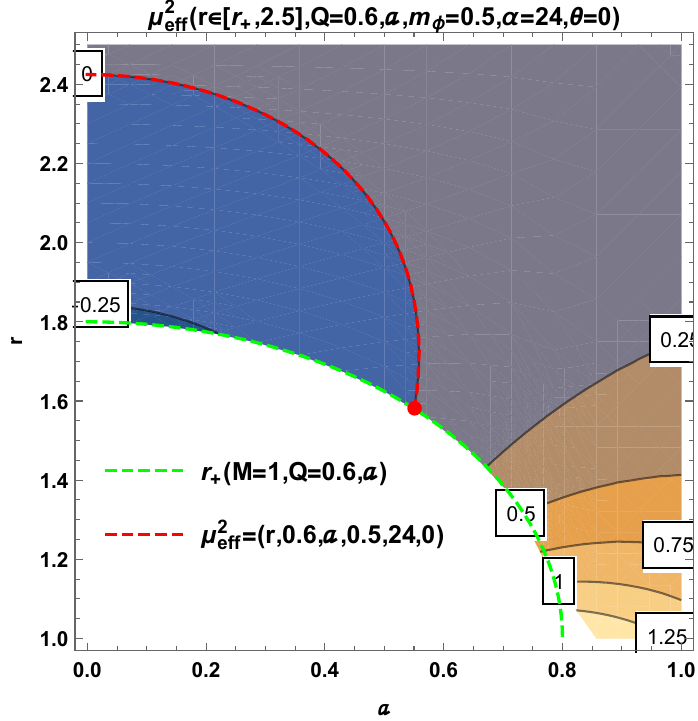}}
  \hfill%
\subfigure[~ $Q=0.6$,~$\theta=\frac{\pi}{3}$ ]{  \includegraphics[width=0.3\textwidth]{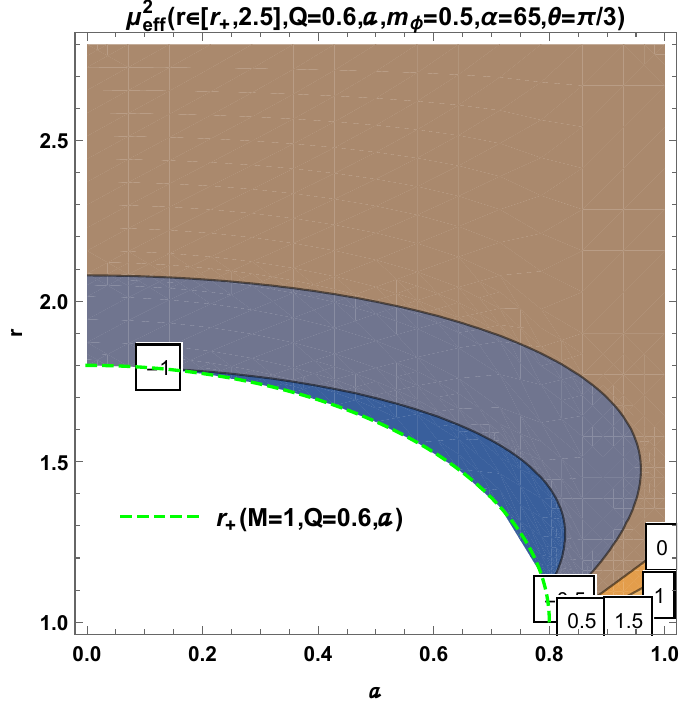}}
 \hfill%
\subfigure[~ $Q=0.6$,~$\theta=\frac{0.9\pi}{2}$]{  \includegraphics[width=0.3\textwidth]{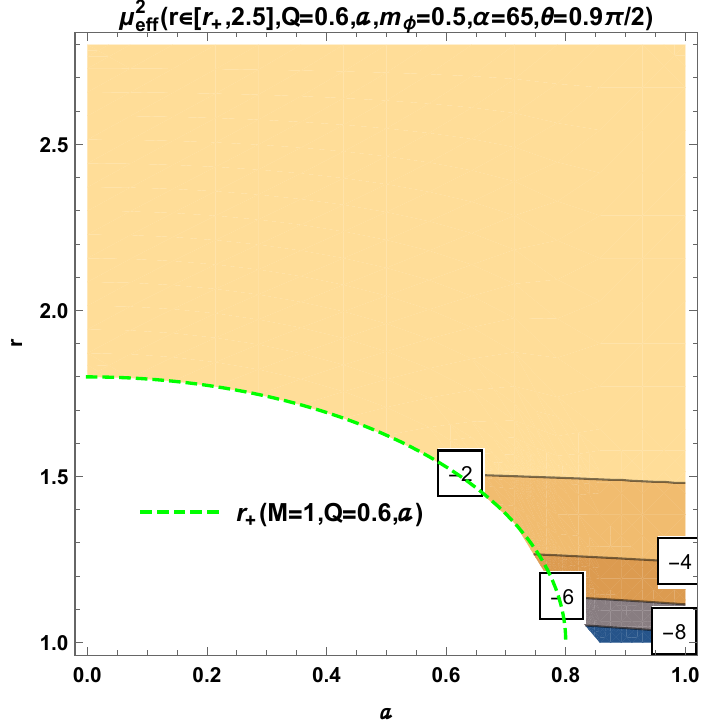}}
\caption{Graphs for showing the sign change  of $\mu_{\rm eff}^2(r\in[r_+,2.5],Q=0.6,a\in[0,1],m_\phi=0.5,\alpha=24,\theta)$ as functions of $r$ and spin $a$ with three different $\theta$. {{Here,}
 $\mu_{\rm eff}^2=0$ {represents its zero value (red-dashed curve) and} $r_+(M=1,Q=0.6,a\in[0,0.8])$ denotes the horizon radius (green-dashed curve). The different shaded regions correspond to their different $\mu_{\rm eff}^2$ values between boundary curves. }  
(\textbf{a}) $\theta=0$. One finds that the  negative region is given by $0<a<a_o$(=0.5509, red dot).
(\textbf{b}) $\theta=\frac{\pi}{3}$. One finds the whole negative region in the near-horizon.(\textbf{c}) $\theta=\frac{0.9\pi}{2}$. The whole negative region is found in the near-horizon.}\label{fig1}
\end{figure}

\begin{figure}[H]
\subfigure[~ $Q=0.4$,~$\theta=0$ ]{  \includegraphics[width=0.3\textwidth]{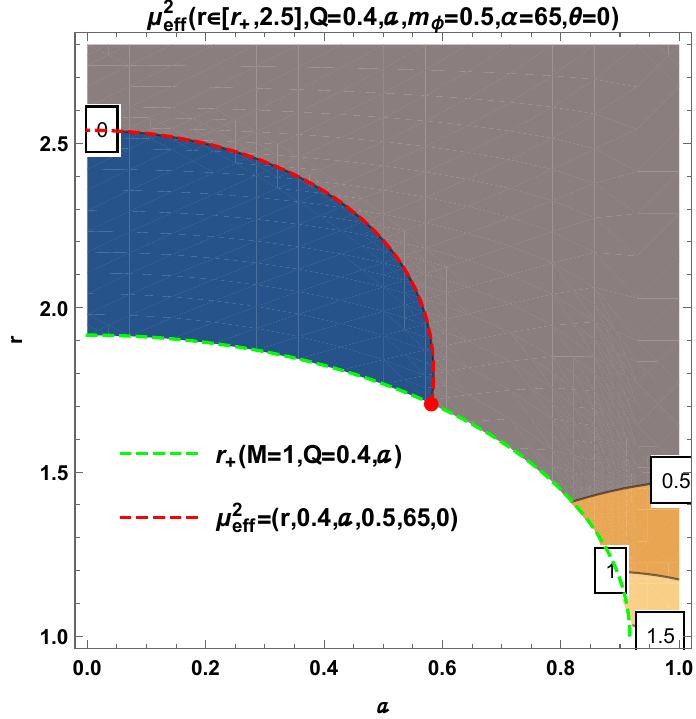}}
  \hfill%
\subfigure[~ $Q=0.4$,~$\theta=\frac{\pi}{3}$ ]{  \includegraphics[width=0.3\textwidth]{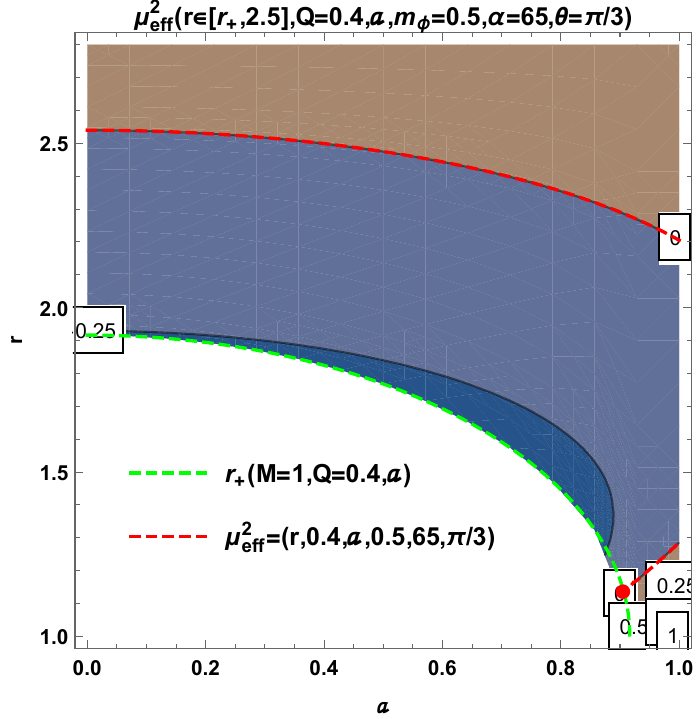}}
 \hfill%
\subfigure[~ $Q=0.4$,~ $\theta=\frac{0.9\pi}{2}$]{  \includegraphics[width=0.3\textwidth]{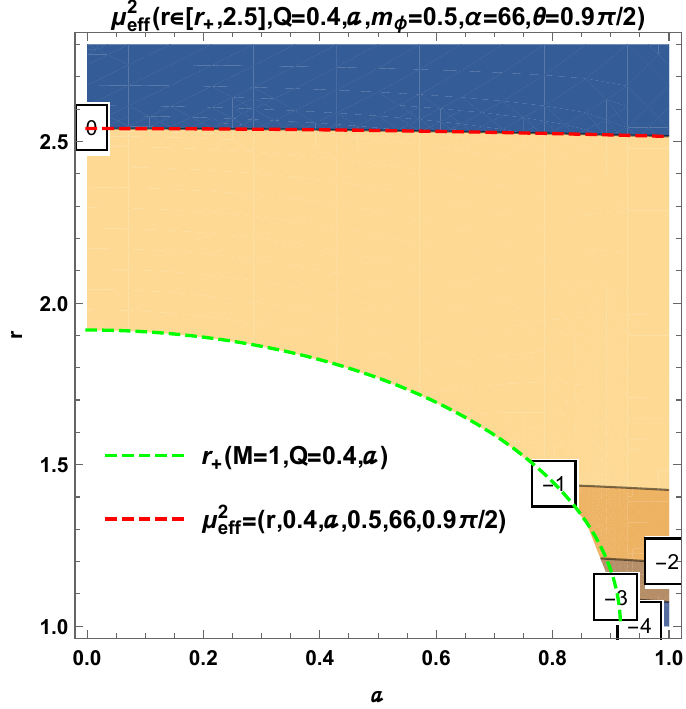}}
\caption{Graphs for showing the sign change  of $\mu_{\rm eff}^2(r\in[r_+,2.5],Q=0.4,a\in[0,1],m_\phi=0.5,\alpha=65,\theta)$ as functions of $r$ and spin $a$ with three different $\theta$. Here, $\mu_{\rm eff}^2=0$ represents its zero curve (red-dashed) and $r_+(M=1,Q=0.4,a\in[0,0.917])$ denotes the horizon radius (green-dashed).  
(\textbf{a}) $\theta=0$. One finds that the  negative region is allowed for $0<a<a_o$(=0.5818, red dot).
(\textbf{b}) $\theta=\frac{\pi}{3}$.  One finds that the  negative region is given by $0<a<a_o$(=0.9062, red dot).
(\textbf{c}) $\theta=\frac{0.9\pi}{2}$. One finds the whole negative region in the near-horizon.}\label{fig2}
\end{figure}

One observes from Figure~\ref{fig1}  with $Q=0.6$ and  $\alpha=24$ the appearance of a negative region for $\mu^2_{\rm eff}|_{\theta=0}$   in the near-horizon: $0<a<a_o(=0.5509)$.
However, for $\theta=\frac{\pi}{3}, \frac{0.9 \pi}{2}$, there are  no onset spins.
We  observe from Figure~\ref{fig2}  with $Q=0.4$ and  $\alpha=65$ the appearance of negative regions for $\mu^2_{\rm eff}|_{\theta=0,\frac{\pi}{3}}$  in the near-horizon: $0<a<a_o(=0.5818)$ and $0<a<a_o(=0.9062)$. 
However, for $\theta= \frac{0.9 \pi}{2}$, there is  no onset spin.
This implies that as a whole, there is no spin-bound on the onset of scalarization with positive coupling parameter $\alpha$ because the contribution of $\mu^2_{\rm eff}$ around ($\theta=\pi/2$) is dominant negatively.

\vspace{-9pt}

\section{Spin-Charge-Induced Scalarization }\label{3s}

We adopt the numerical method to solve the linearized scalar  equation. For this purpose, the Kerr azimuthal coordinate $\varphi*$ and the tortoise coordinate $x$ are introduced as 
\begin{eqnarray}
  d\varphi^* &=& d\varphi +\frac{a}{\Delta}dr, \nonumber\\
  dx &=& \frac{r^2+a^2}{\Delta}dr. \label{new-coor}
\end{eqnarray}

Taking into account the axial symmetry of the KN background in Equation~\eqref{KN-sol}, the scalar perturbation could be decomposed as
\begin{equation}
\delta\phi(t,x,\theta,\varphi^*)=\sum_{m}\delta\phi(t,x,\theta)e^ {im\varphi^*} \label{s-dec}
\end{equation}
with $m$ an azimuthal number. Substituting \eqref{s-dec} into the linearized scalar Equation (\ref{per-eq}), we have the $(2+1)$-dimensional evolution   equation
\begin{eqnarray}
  &&\left[-(r^2+a^2)^2+\Delta a^2\sin^2\theta\right]\partial^2_t\delta\phi+(r^2+a^2)^2\partial^2_x\delta\phi
  +\Delta\partial^2_\theta\delta\phi\nonumber\\
  &&-2ima{(r^2+a^2-\Delta)}\partial_t\delta\phi+2\left[r\Delta+ima(r^2+a^2)\right]\partial_x\delta\phi+\Delta\cot{\theta}\partial_\theta\delta\phi \nonumber\\
  &&+\Delta\Big[-\mu^2_{\rm eff}(r^2+a^2\cos^2\theta)+\frac{m^2}{\sin^2\theta}\Big]\delta\phi= 0. \label{mscalar-eq2}
\end{eqnarray}
where the tortoise coordinate $x\in(-\infty,\infty)$ covers the whole region that is accessible to an observer
located outside the outer horizon, while one confines the semi-infinite region  $r\in[r_+,\infty)$ when using the coordinate $r$.

{{Considering}   the massive linearized scalar field, it may lead to instability or errors in numerical computations if directly applying the hyperboloidal slicing method. Here, we adopt the direct 2 + 1 time evolution method \cite{Doneva:2020kfv,Zhang:2021btn} to calculate the linearized massive scalar perturbation on the KN black hole background.}
The details on the code implementation were described  in Ref. \cite{Lai:2022spn}.
Here,  we wish to mention  the important things on how to solve Equation (\ref{mscalar-eq2}).
The derivatives in $x$- and $\theta$-directions are approximated by making use of a finite difference method,
while the time evolution is carried out by adopting the fourth-order Runge--Kutta integrator. 
Moreover, we impose physical boundary conditions: an ingoing wave at the horizon and an outgoing wave at infinity. 
In practical calculations, one has to truncate the infinite radial computational domain to a finite range and put boundary conditions at the outer edges.  It induces  inevitably resulting spurious wave reflections from the outer edges.  To overcome this “outer-boundary problem”, one can  push the outer edges to very large values so that the spurious reflections will not affect the observed signal for a sufficiently long evolution time. 
At the poles of $\theta=0$ and $\pi$, we impose the physical boundary condition of  $\Psi|_{\theta=0,\pi}=0$ for $m\neq0$, while $\partial_\theta\Psi|_{\theta=0,\pi}=0$ is imposed  for $m=0$.  

Importantly,  the initial data of the scalar perturbation are chosen to be  a Gaussian distribution localized outside the horizon  with time symmetry as  
\begin{subequations}
\begin{align}
\Psi(t=0,x,\theta) &\sim P^l_m(\theta) e^{-\frac{(x-x_c)^2}{2\sigma^2}}, \label{eq:initialPsi} \\
\Pi(t=0,x,\theta) &= 0. \label{eq:initialPi}
\end{align}
\end{subequations}  
where $P^l_m(\theta)$ denotes the associated Legendre polynomials,   $x_c$ represents a location of the center, and $\sigma$ indicates  a width of the distribution. 
It is worthy to point out that although  there is one initial mode with a specified $l$ only,
other $l$ modes with the same index $m$ will be activated during the evolution.  Hence, it is necessary to consider the influence of $l$ and $m$ of the initial perturbation. However, we note that the $l=m$ mode will have a dominant contribution at late times. To our knowledge,  similar phenomena have  occurred in other theories. Hereafter,  we choose the initial mode $\Psi$ with $l=m=0$ for simplicity.

\section{Numerical Results}\label{4s}

In this section,  we describe  schematic  pictures concerning the influences of charge $Q$, spin $a$,  coupling parameter $\alpha$, and  scalar mass $m_\phi$ on spin-charge-induced  scalarization  by  solving   the (2 + 1)-dimensional equation for  a massive scalar.
Before we proceed, we note that the Kerr spacetime is not spherically symmetric except when $a=0$, so the mode-coupling  phenomenon occurs~\cite{Cunha:2019dwb,Collodel:2019kkx,Herdeiro:2020wei}: a pure initial $l$-multipole will excite other multipoles with the same $m$ as it evolves. Taking into account this phenomenon and for simplicity,  we  consider  $l=m=0$ mode for the KN black hole background.  

\begin{figure}[H]
 \includegraphics[width=0.65\textwidth]{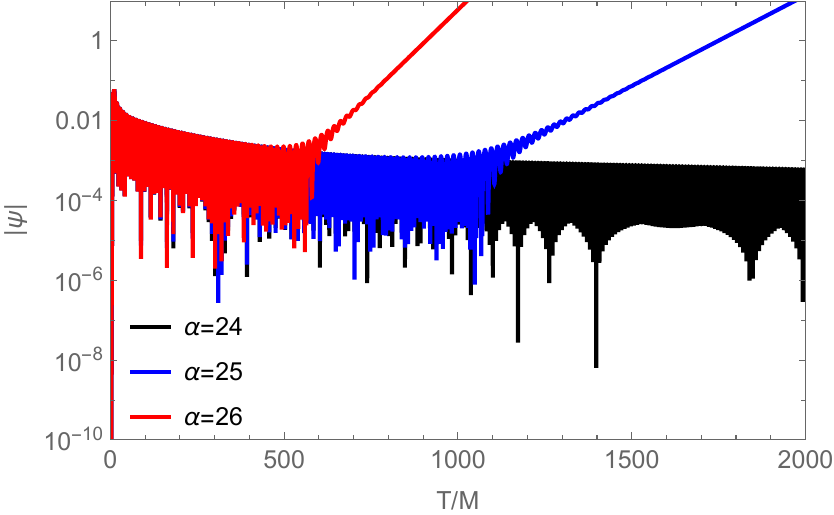}
\caption{Time-domain profile of  $|\Psi|$ for $m_\phi=0.5$, $Q=0.6$, and
$a=0.4$ with three coupling parameters  $\alpha=24,25, 26$. Here, $\alpha=24.4$ represents the threshold value ($\alpha_{\rm th}$).  Hence, $\alpha=25$ and 26 are unstable modes, while $\alpha=24$ is a stable mode.   }\label{fig3}
\end{figure}

\begin{figure}[H]
\includegraphics[width=0.65\textwidth]{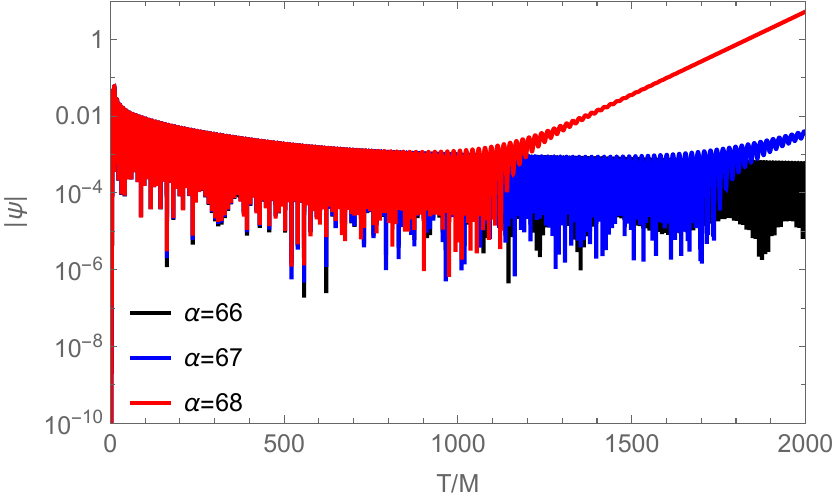}
\caption{Time-domain profile of $|\Psi|$ for $m_\phi=0.5$, $Q=0.4$, and $a=0.5$ with three couplings  $\alpha=66,67, 68$. Here, $\alpha= 66.7$ represents the threshold value ($\alpha_{\rm th}$). }\label{fig4}
\end{figure}

Considering different values of coupling parameter $\alpha$, the time domain profile of the $l=m=0$-scalar mode $|\Psi|$ is shown in Figure~\ref{fig3}. 
We note that  $\alpha_{\rm th}=24.4$  corresponds to  the threshold (marginal) evolution of tachyonic instability for $m_\phi=0.5,~Q=0.6$, and $a=0.4$.  From this figure, it is easily found  that tachyonic instability is  triggered once the coupling constant $\alpha$ exceeds $\alpha_{\rm th}=24.4$. Similarly, Figure~\ref{fig4} indicates  time-domain profiles  of the scalar mode $|\Psi|$ with $m_\phi=0.5,~Q=0.4$, and $a=0.5$ for three  couplings $\alpha=66,67$, and $68$.
Here, its threshold value is given by $\alpha_{\rm th}=66.7$. We expect to find a similar time-domain profile of $\Psi$ for the $Q=0.2$ case. {{Actually,}
 we note that  $\alpha_{\rm th}=24.4,66.7$ can be predicted when finding  its negative region of scalar potentials around the Kerr--Newman black hole.  Qualitatively, the presence  of the negative region for $\mu_{\rm eff}$ in Figures~\ref{fig1} ($\alpha=24$) and \ref{fig2} ($\alpha=65$) implies that  $\alpha_{\rm th}$ might take its values around $=24$ and 65. However,  the precise value of $\alpha_{\rm th}$ could be determined  by observing  the time-domain profile $|\Psi|$.  }

\vspace{3pt}

\vspace{-3pt}

\begin{figure}[H]
  \subfigure[~$m_\phi=0$]{\includegraphics[width=0.3\textwidth]{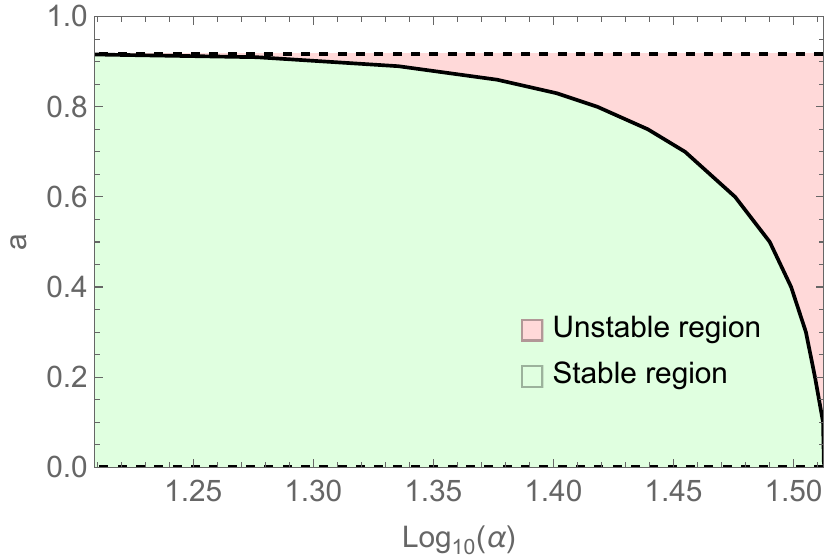}}
  \hfill%
 \subfigure[~$m_\phi=0.5$]{ \includegraphics[width=0.3\textwidth]{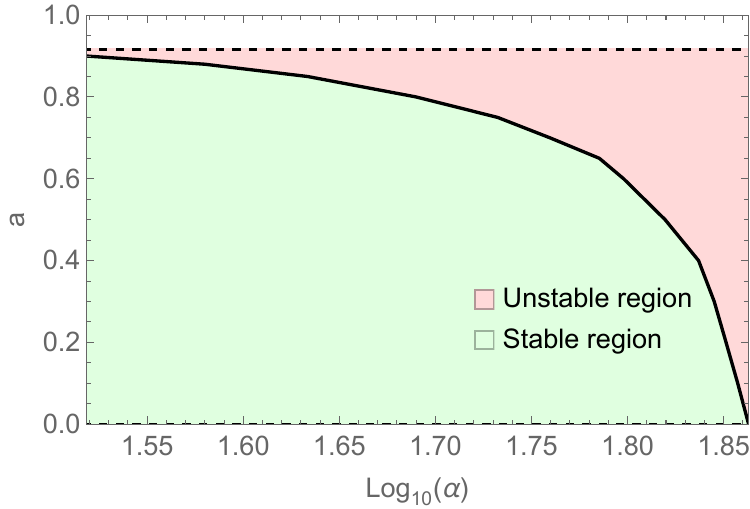}}
  \hfill%
\subfigure[~$m_\phi=1$]{  \includegraphics[width=0.3\textwidth]{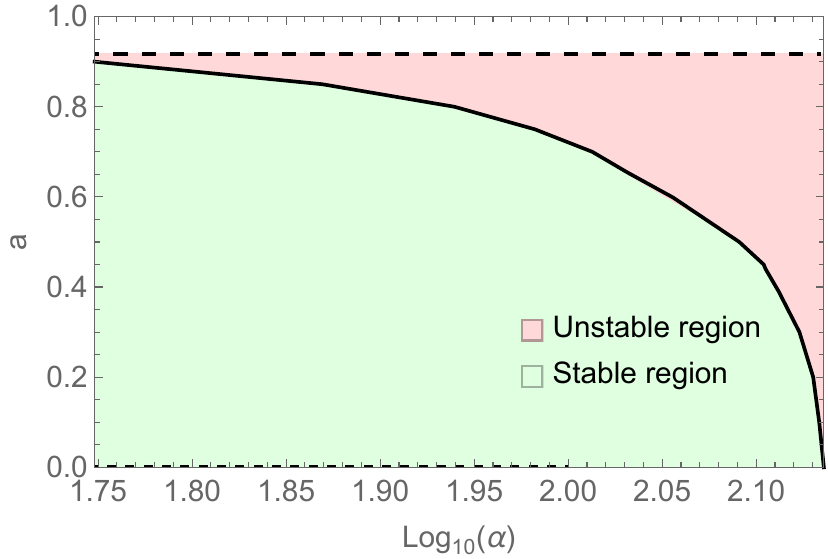}}
\caption{Threshold curves $\log_{10}\alpha_{\rm th}(a)$ for $Q=0.4$ with different scalar mass $m_\phi=0,~0.5,~1$.  All curves denote the boundaries between stable (lower) and unstable (upper) region. Black solid lines
represent the existence condition [$a^2\le 1-Q^2(=0.84)$] of the outer horizon.}\label{fig5}
\end{figure}

\begin{figure}[H]

   \subfigure[~$m_\phi=0$]{ \includegraphics[width=0.3\textwidth]{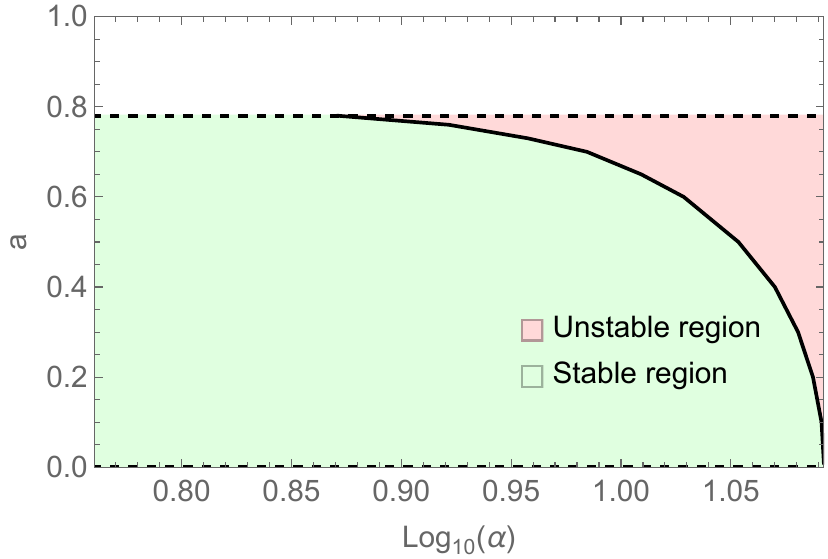}}
  \hfill%
  \subfigure[~$m_\phi=0.5$]{  \includegraphics[width=0.3\textwidth]{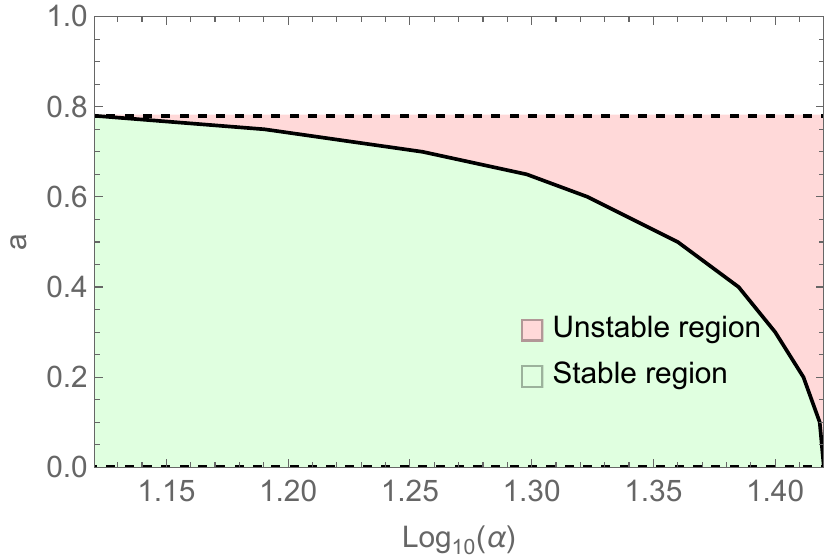}}
  \hfill%
   \subfigure[~$m_\phi=1$]{ \includegraphics[width=0.3\textwidth]{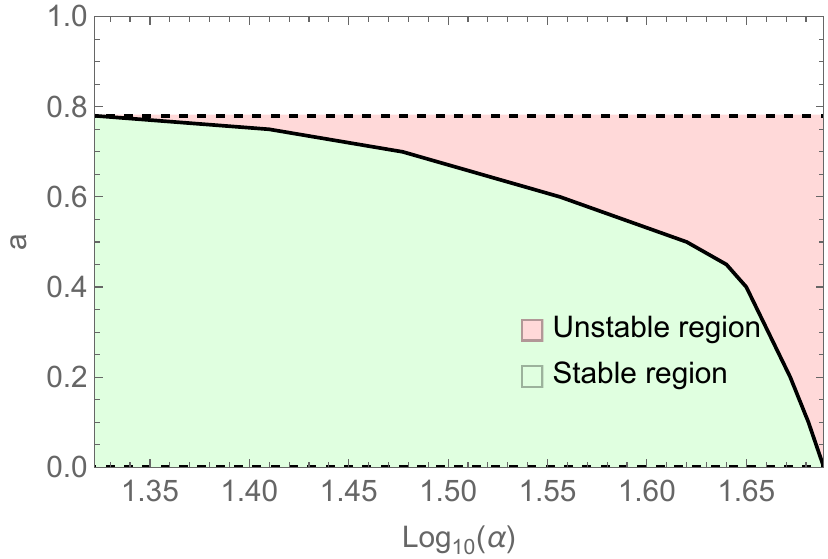}}
\caption{Threshold curves $\log_{10}\alpha_{\rm th}(a)$ for $Q=0.6$ with three scalar mass $m_\phi=0,~0.5,~1$.  All curves denote the boundaries between stable (lower) and unstable (upper) region. Black solid lines
represent the existence condition [$a^2 \le 1-Q^2(=0.64)$] of the outer horizon.}\label{fig6}
\end{figure}

Importantly, we obtain  the threshold  curve [$\log_{10}\alpha_{\rm th}(a)$] with different charge $Q$ 
and scalar mass   $m_{\phi}$   by making use of  the  $l=m=0$-mode $\Psi$.
A more clear picture for the influence of two parameters  on the threshold curve is depicted  in Figures~\ref{fig5} and \ref{fig6}. The threshold curve $\log_{10}\alpha_{\rm th}(a)$ describes the boundary between stable and unstable KN black holes.
The allowed region for spin $a$ is determined by  its existence condition. The unstable region  for scalarized KN black holes  depends on the scalar mass.

\begin{figure}[H]
  \includegraphics[width=0.3\textwidth]{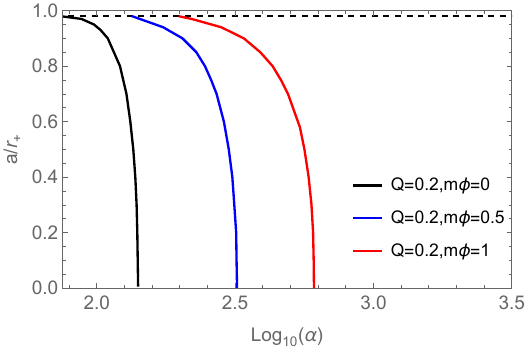}
  \hfill%
  \includegraphics[width=0.3\textwidth]{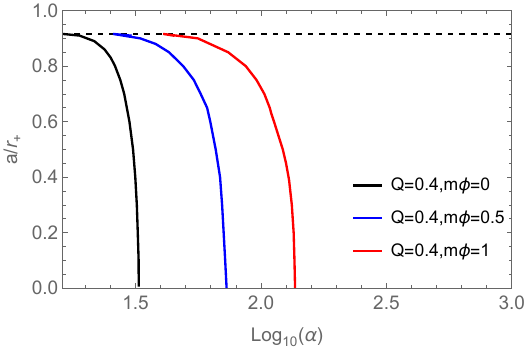}
  \hfill%
  \includegraphics[width=0.3\textwidth]{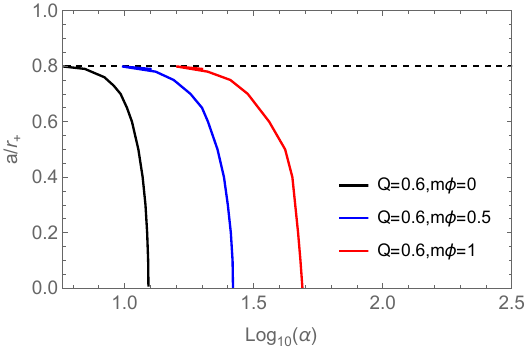}
\caption{Threshold curves $\log_{10}\alpha_{\rm th}(a)$ for three  charges  $Q=0.2,~0.4,~0.6$ with three  scalar masses $m_\phi=0,~0.5,~1$.  All  curves denote the boundaries between the stable  and unstable  region. Black solid lines determine  the existence condition $a^2(Q) \le 1-Q^2$ of the outer horizon.}\label{fig7}
\end{figure}

The termination point of each curve determined by $\log_{10}\alpha_{\rm th}(a=0,Q,m_\phi)$  increases as the scalar mass increases. 
For $m_\phi=0$, the scalar  becomes massless, and its unstable region  is the largest one when comparing with $m_\phi\not=0$. Figure~\ref{fig7} shows that for given $Q$, three threshold curves are located from left to right as $m_\phi$ increases. The position of the termination point increases as the scalar mass increases.   This confirms  clearly that the presence of scalar mass  suppresses  or  quenches  the tachyonic instability in the KN spacetime~\cite{Doneva:2020kfv,Zhang:2021btn}.  Finally, considering the existence condition of the outer horizon Equation~(\ref{l-b}) with $M=1$, one may  determine  the upper bound of $a$. For $Q=0.2,~0.4,$ and 0.6, the maximum values (upper bounds) of $a$ are given by 0.980, 0.917, and 0.8, respectively (see Figures~\ref{fig1}, \ref{fig2}, \ref{fig5}--\ref{fig7}).

\vspace{-9pt}

\vspace{-5pt}

\section{Conclusions and discussions}\label{5s}


In this work, we have analyzed the spin-charged induced scalarization  of KN black holes described by mass $M=1$, charge $Q$, and spin $a$ in the EMS theory with scalar mass  $m_\phi$
and  positive scalar coupling $\alpha>0$. We found that  there exists  a bound of $0<a<a_o$ with  onset spin  $a_o(Q,m_\phi,\alpha)$ for  the negative region signaling  instability  by analyzing the effective scalar mass term $\mu^2_{\rm eff}$ in the $\theta$-direction.
However, it implies that  as a whole, there is no spin-bound on the onset of scalarization with positive coupling parameter $\alpha$ because  contribution of $\mu^2_{\rm eff}$ around ($\theta=\pi/2$) is dominant negatively. 
We adopted the the fourth-order Runge-Kutta integrator to solve 
the $(2+1)$-dimensional  evolution equation.
 We have obtained the threshold curves  [$\log_{10}\alpha_{\rm th}(a,Q,m_\phi)$] to describe 
boundaries  between stable (lower region) and unstable (upper region) KN black holes. 
Also, all threshold curves increase  as $a$ decreases with upper bound determined by $a^2\le 1-Q^2$. The unstable region is allowed  only  for constructing scalarized KN black holes. 

Finally, the computation including  nonlinear
effects is  expected to quench the tachyonic  instability and leads to constructing  scalarized KN black holes. Hence, a promising  direction is to construct the scalarized KN black holes in the EMS theory with  positive scalar coupling and  mass term.

 \vspace{1cm}

{\bf Acknowledgments}

This work is supported by the National Natural Science Foundation of China (NSFC) with Grant, Nos.~12365009, 12305064 and 12205123.

\end{document}